\documentclass[12pt]{article}
\usepackage{RR}
\usepackage[T1]{fontenc}
\usepackage{a4wide}
\usepackage{url}
\usepackage{comment}
\usepackage{aeguill}
\usepackage{amsmath,amssymb,amsthm}
\newtheorem{lemma}{Lemma}
\newtheorem{coro}{Corollary}
\newtheorem{theorem}{Theorem}
\newtheorem{definition}{Definition}

\def\P{{\mathcal P}}
\def\N{{\mathcal N}}
\def\Z{{\mathbb Z}}
\def\disc{\mathop{\mathrm{disc\,}\relax}}
\def\lc{\mathop{\mathrm {lc}\,}}
\RRdate{Janvier 2007}
\RRauthor{Howard Cheng\thanks{Department of Mathematics and Computer Science, University of Lethbridge, {\tt howard.cheng@uleth.ca}} \and Guillaume Hanrot \and Emmanuel Thom\'e \and
Eugene Zima\thanks{Wilfrid Laurier University, {\tt ezima@wlu.ca}} \and Paul Zimmermann}
\authorhead{H.~Cheng et {\em al.}}
\RRtitle{\'Evaluation de certaines constantes hyperg\'eom\'etriques 
efficace en temps et en espace}
\RRetitle{Time- and Space-Efficient Evaluation of Some Hypergeometric
Constants}
\RRabstract{
  The currently best known algorithms for the numerical evaluation of
  hypergeometric constants such as $\zeta(3)$ to $d$ decimal digits
  have time complexity $O(M(d) \log^2 d)$ and space complexity of $O(d \log d)$
  or $O(d)$.
  Following work from Cheng, Gergel, Kim and Zima, we
  present a new algorithm with the same asymptotic complexity, but
  more efficient in practice.
  Our implementation of this algorithm improves slightly over
  existing programs for the computation of $\pi$,
  and we announce a new record of 2 billion digits for $\zeta(3)$.
}
\RRresume{  
Les meilleurs algorithmes actuels pour l'\'evaluation num\'erique de 
constantes hyperg\'eom\'etriques comme  $\zeta(3)$ avec $d$ chiffres
de pr\'ecision ont une complexit\'e en temps $O(M(d) \log^2 d)$ 
et en espace $O(d \log d)$ ou $O(d)$. Dans la lign\'ee de travaux
de Cheng, Gergel, Kim and Zima, nous pr\'esentons un nouvel
algorithme de m\^eme complexit\'e asymptotique, mais plus efficace
en pratique. Notre impl\'ementation de cet algorithme est 
plus efficace que les codes existants pour les calculs de $\pi$,
et nous annon\c{c}ons un nouveau record de 2 milliards de chiffres
d\'ecimaux for $\zeta(3)$.
}

\RRkeyword{Hypergeometric constants, binary splitting, sieve}
\RRmotcle{Constantes hyperg\'eom\'etriques, scindage binaire, crible}
\RRprojet{Cacao}
\RRtheme{\THSym}
\URLorraine

\begin{document}
\makeRR

\section{Introduction}
In this article, we are interested in the high-precision evaluation of
constants defined by hypergeometric series of the form
\begin{equation}
  \sum_{n=0}^{\infty} a(n) \prod_{i=0}^{n-1} \frac{p(i)}{q(i)},
  \label{hypergeom}
\end{equation}
where $a$, $p$ and $q$ are polynomials with integer coefficients.  We
shall also assume, without loss of generality, that $p$ and $q$ are
coprime, have no nonnegative integer as a zero and that $p(n)/q(n)$ tends
to a constant $0 < c < 1$ when $n$ goes to infinity.

Under those assumptions, the series converges, and we can compute an 
approximation to the constant by truncating the series, 
i.e., by computing
\begin{equation}
  \label{eq:hypergeomsum}
  \sum_{n=0}^{N-1} a(n) \prod_{i=0}^{n-1} \frac{p(i)}{q(i)}
\end{equation}
for an appropriately chosen $N = \Theta(d)$ with $d$ being the number of
decimal digits desired.
The high-precision
evaluation of elementary functions and other constants ---including the
exponential function, logarithms, trigonometric functions, and
constants such as $\pi$ or Ap\'ery's constant
$\zeta(3)$--- is commonly carried out by evaluating
such series~\cite{GoSe01,HaPa98}.  For example, we have
\begin{equation} \label{pi3e}
\frac{1}{\pi} = 12 \sum_{n=0}^{\infty} (-1)^n
\frac{545140134n+13591409}{640320^{3n+3/2}}
\frac{(6n)!}{(3n)! n!^3}
\end{equation}

or
\begin{equation} \label{zeta3}
2 \zeta(3) = \sum_{n=0}^{\infty} (-1)^n (205n^2+250n+77)
\frac{(n+1)!^5
  n!^5}{(2n+2)!^5}.
\end{equation}

Assuming that $q(n)$ has size $O(\log n)$, the special form of the series
(\ref{eq:hypergeomsum}) implies that the common denominator
$\prod_{i=0}^{N-2} q(i)$ has a relatively small size of $O(N \log N)$.  An
approach commonly known as ``binary splitting'' has been independently
discovered and used by many authors in such
computations~\cite{Be04,BoBo87,BoBrCr00,Br76,Go90,GoSe01,HaPa98,Karatsuba97}.
In binary splitting, the use of fast integer multiplication yields a total time
complexity of $O(M(d\log d)\log d) = O(M(d) \log^2 d)$, where $M(t) = O(t
\log t \log \log t)$ is the complexity of multiplication of two
$t$-bit integers~\cite{ScSt71}.  The $O(d \log d)$ space complexity of
the algorithm is the same as the size of the computed numerator and
denominator.

The numerator and denominator computed by the binary splitting approach
typically have large common factors.  For example, it was shown that in
the computation of 640,000 digits of $\zeta(3)$, the size of the reduced
numerator and denominator is only 14\% of the size of the computed
numerator and denominator. This suggests possible improvements of the
method, by avoiding the unneeded computation of the common divisor
between the numerator and denominator.  Several approaches have already
been taken in that direction. In particular,
\cite{ChZi00} suggests to use a partially factored form for the computed
quantities, in order to efficiently identify and remove common factors,
and \cite{Xue02} goes further by explicitly constructing the common divisor
and dividing out the numerator and denominator.

The present work builds on top of this strategy and uses a fully factored
form in the binary splitting process. We show that the fully factored
form yields a time complexity of $O(M(d) \log^2 d)$, and space complexity
$O(d)$. This matches the complexity of the standard approaches, but
provides a practical speedup confirmed by experiments. Our method
appears to be noticeably faster than other optimized binary splitting
implementations aimed at the computation of digits of $\pi$ or other
constants. We also show in this article that the exact set of series that
are amenable to efficient computation using the fully factored form is
characterized by a simple criterion: only the series where $p(n)$ and
$q(n)$ are products of linear factors exhibit the large common factor
that was observed in the computation of $\zeta(3)$. Therefore our
attention is restricted to that case.
\medskip

This article is organized as follows. Section~\ref{sec:algo-classic}
recalls the binary splitting algorithm, and reviews the different
approaches for improving the practical efficiency of the method.
Section~\ref{sec:fraction} examines in detail the size of the reduced
fraction computed by the binary splitting algorithm.
Section~\ref{sec:our-algo} presents the alternative of using a fully
factored form in the binary splitting approach. In
Section~\ref{sec:analysis}, the analysis of the algorithm is performed.
Section~\ref{sec:experiments} concludes with experimental data, and a
comparison with other programs.

\section{The binary splitting approach and its variants}
\label{sec:algo-classic}
We give a brief description of the binary splitting approach here,
following the notations from \cite{ChZi00}.

Our approximation to the constant to be evaluated can be written $S(0,
N)$ where for $0 \leq n_1 \leq n_2$ we define
\[ S(n_1,n_2) = \sum_{n=n_1}^{n_2-1} a(n) \frac{p(n_1) \cdots p(n-1)}
                                                    {q(n_1) \cdots
q(n-1)}. \]
Letting $P(n_1,n_2) = \prod_{n=n_1}^{n_2-1} p(n)$,
$Q(n_1,n_2) = \prod_{n=n_1}^{n_2-1} q(n)$,
and $T(n_1,n_2) = S(n_1,n_2) Q(n_1,n_2)$, we have for $n_1 < m < n_2$,
with $T(n, n+1) = a(n) p(n)$:
\begin{align*}
P(n_1,n_2) &= P(n_1,m) P(m,n_2) \\
Q(n_1,n_2) &= Q(n_1,m) Q(m,n_2) \\
T(n_1,n_2) &= T(n_1,m) Q(m,n_2) + P(n_1,m) T(m,n_2).\mskip-60mu
\end{align*}
This leads to a recursive algorithm to evaluate
$T(0, N)$ and $Q(0, N)$, which corresponds to the
evaluation of a product tree~\cite{Be04}.
One then deduces $S(0, N)$ by a division. 

Since $p(n)/q(n)$ tends to $c$, the tail $S(N,\infty)$ of the series is
bounded by $O(c^N)$.  Therefore, to compute the constant $S(0,\infty)$
with error $\epsilon$, we need $N=\frac{\log\epsilon} {\log c} + O(1)$
terms: the number of terms is proportional to the number $d$ of digits
of accuracy desired.  The corresponding product tree has height $\log N$,
where the leaves have $O(\log N)$ bits and hence the root has $O(N \log
N)$ bits.  The total evaluation of the truncated series costs
$O(M(d \log d) \log d) = O(M(d) \log^2 d)$ with the best-known
multiplication algorithms.

Although some constants such as $\pi$ and $\log 2$ can be computed to $d$
digits with bit complexity of $O(M(d) \log d)$ using the Arithmetic
Geometric Mean (AGM) \cite{BoBo87},
the $O(M(d) \log^2 d)$ binary splitting algorithm is still competitive
up to billions of digits. For example, D.~ V.~and G.~V.~Chudnovsky
held the $\pi$ record using Formula (\ref{pi3e}) with 8 billion
digits in 1996 \cite{GoSe01}.

\subsection{Improvements of the binary splitting method}

As mentioned earlier, the binary splitting method suffers
from the drawback that the fraction $T/Q$ has size $O(d\log
d)$, while an accuracy of only $d$ digits is required. In~\cite{GoSe01},
the authors circumvent this problem by limiting the precision of the
intermediate results to $O(d)$ digits.  This is used by the
\texttt{PiFast} program~\cite{GoSe01} and results in the same time
complexity as the binary splitting method but a reduced space complexity
of $O(d)$. This truncation, however, implies that the exact reduced
fraction is not computed, so that it is not easy to extend the
computation to higher precision using results already computed. Further,
the truncation only operates on the top levels of the
computation tree, since below a depth of order $O(\log\log d)$, the
computed integers have size $O(d)$ anyway. Below this depth, the
computations performed by the \texttt{PiFast} program are expected to be
exactly the same as in the classical algorithm above.
\smallskip

Since in the course of the computation of digits of $\zeta(3)$, $T$ and
$Q$ have been found to share a large number of common factors,
Cheng and Zima~\cite{ChZi00} worked towards efficiently removing some of
these factors from the computation.  For this purpose, a partially
factored representation was introduced in the binary
splitting process.  Subsequently, Cheng, Gergel, Kim, and
Zima~\cite{ChGeKiZi05} applied modular computation and rational number
reconstruction to obtain the reduced fraction.  If the reduced numerator
and denominator have size $O(d)$, the resulting algorithm has a space
complexity of $O(d)$ and the same time complexity as binary splitting.
By carefully analyzing the prime divisors of the numerator and
denominator of (\ref{zeta3}), it was shown in \cite{ChGeKiZi05} that the
size of the reduced fraction for $\zeta(3)$ is $O(d)$;  it was noted that
the analysis was in fact related to using the partially factored
representation with all possible prime factors in the binary splitting
process.  However, it was not practical to use so many
primes in the partially factored representation because it was expensive
to convert from standard representation by factoring.  Additionally
practicality of the algorithm depends on the availability of the
implementation of the asymptotically fast rational reconstruction
algorithm (for example, see \cite{pan-wang:2003}).
\smallskip

We also mention the \texttt{gmp-chudnovsky} program~\cite{Xue02}, which
uses the binary splitting method to compute digits of $\pi$ using
Formula~(\ref{pi3e}).  Two modifications are made to the classical method
described above. First, integers $P(n_1,n_2)$ and $Q(n_1,n_2)$ are handled
together with
their complete factorization. This makes it possible to quickly compute
the gcd of $P(n_1,m)$ and $Q(m,n_2)$ by merely comparing the
factorizations. Afterwards, the gcd is divided from both numbers. The
fraction $T/Q$ is therefore reduced. It should be noted that
\texttt{gmp-chudnovsky} still works with expanded integers $P$, $Q$,
and $T$ (albeit reduced).

The second specificity of the \texttt{gmp-chudnovsky} program lies in the
way the leaves $p(n)$ and $q(n)$ are computed. Since the factorization of
these numbers is sought, an optimized sieve table is built.
Formula (\ref{pi3e}) implies that the integers to be factored are bounded by
$6N$, where $N$ is the number of computed terms. A table of
$\left\lfloor\frac{6N}2\right\rfloor$ entries is built, with the $i$-th
cell containing information on the smallest prime divisor of $2i+1$, its
multiplicity, and the integer $j$ such that $2j+1$ is the cofactor. Such
a table can be computed very efficiently using a modified Eratosthenes'
sieve. This represents a tiny part of the total computing time.
Unfortunately, this sieve table is also an impediment to large scale
computations, in that it has a space complexity of
$O(d\log d)$.


\section{Size of the reduced fraction}
\label{sec:fraction}

Cheng, Gergel, Kim and Zima showed in \cite{ChGeKiZi05} that for Formula
(\ref{zeta3}) giving $\zeta(3)$, when removing common factors between $T$
and $Q$, the reduced fraction $\hat{T}/\hat{Q}$ has size $O(d)$ only.  We
show here that this fact happens for a large class of hypergeometric
constants.

Understanding when the size of the fraction reduces
to $O(d)$ is closely linked to a study of the prime divisors of the values 
$p(i)$ and $q(i)$. Indeed, the fraction being significantly smaller
than its expected $O(d\log d)$ size means that there are large cancellations
at many primes; it thus means that the primes occurring in 
$\prod_{i=n_1}^{n_2-1} p(i)$ and $\prod_{i=n_1}^{n_2-1} q(i)$ are mostly
the same, and with the same multiplicities. 

We first notice that since $p(n)/q(n)$ tends to $c > 0$ when $n
\rightarrow \infty$, this implies that $p$ and $q$ have the same degree.
For a polynomial $p$, we use the notation $\lc(p)$ and $\disc(p)$ to
denote the leading coefficient and the discriminant of $p$,
respectively.  If $p$ is an irreducible polynomial and $\ell$ a prime
(or prime power) coprime to $\Delta(p) := \lc(p) \disc(p)$, we shall
denote $\rho_\ell(p)$ the number of roots of $p$ modulo $\ell$. If $p
= \prod_{i=1}^k p_i^{e_i}$, and $\ell$ is coprime to $\Delta(p) := 
\prod_{i=1}^k \Delta(p_i)$, we shall define 
$\rho_\ell(p) = \sum_{i=1}^k e_i \rho_\ell (p_i)$, which is still the
number of roots of $p$, counted with multiplicities.

The following lemmata lead to estimates of the $\ell$-valuation and the
size of the quantities $Q(n_1,n_2)$, $T(n_1,n_2)$ and their common
divisors when the summation range $[n_1,n_2]$ grows.

\begin{definition}
Let $\N_{p, \ell}(n_1, n_2)$ be the number of integer
roots of $p(\cdot) \bmod \ell$ in $[n_1, n_2[$:
\[ \N_{p, \ell}(n_1, n_2) := \# \{ x \in [n_1, n_2[ / p(x) = 0 \bmod \ell\}. \]
\end{definition}

\begin{lemma}\label{counting-basic}
Let $p$ be a polynomial, and $\ell$ a prime (or prime power) coprime to
$\Delta(p)$. Then,
$$\N_{p, \ell}(n_1, n_2) = \frac{\rho_\ell(p)}{\ell} (n_2 - n_1) + O(1),$$
where the implied constant in $O(1)$ depends on $p$ only.
\end{lemma}
\begin{proof}
The roots of $p$ modulo $\ell$ in the interval $[n_1, n_2[$ are exactly
integers congruent to one of the $\rho_\ell(p)$ roots of
$p$ in $[0, \ell - 1]$. The Lemma follows, the precise error term being at
most $\rho_\ell(p) \leq \deg p$.
\end{proof}

\begin{definition}
For an integer $m$, let $v_\ell(m)$ be the $\ell$-valuation of $m$, i.e.,
the largest integer $j$ such that $\ell^j$ divides $m$.
\end{definition}
\begin{lemma}\label{lvaluation}
  Let $\ell$ be a prime not dividing $\Delta(p)$. Then,
$$v_\ell(P(n_1, n_2)) = 
\frac{\rho_\ell(p)}{\ell - 1} (n_2 - n_1) + O\left(\frac{\log n_2}{\log \ell} 
\right).$$
\end{lemma}
\begin{proof}
We shall assume without loss of generality that $p$ is irreducible, since by
our definition of $\rho_\ell(p)$, the result in the general case will
follow by linearity.
The $\ell$-valuation of $P(n_1, n_2)$ is exactly
$$v_\ell(P(n_1, n_2)) = 
\sum_{j \geq 1} \N_{p, \ell^j}(n_1, n_2).
$$

Since $\ell$ does not divide $\Delta(p)$, Hensel's Lemma shows that
$\rho_\ell(p) = \rho_{\ell^k}(p)$ for all $k \geq 1$. 

Further, there exists a constant $\gamma$ such that the inequality
$|p(x)| \leq n_2^\gamma$ holds over the interval $[n_1, n_2[$.
We then have $\N_{p, \ell^j}(n_1, n_2) = 0$ for $j > J_p :=
\gamma\frac{\log n_2}{\log\ell}$, and also  $\ell^{-J_p} = O(n_2^{-1})$.
Lemma~\ref{counting-basic} yields for $0 \leq n_1 \leq n_2$:
$$v_\ell(P(n_1, n_2)) = \rho_\ell(p)(n_2 -
n_1)\left(\frac1{\ell-1}\right)+O\left(\frac{\log n_2}{\log
\ell}\right).$$
The statement follows.

\end{proof}

We need to control (though in a rather rough way) what happens for 
primes dividing $\Delta(p)$. The
following weaker lemma is sufficient; its proof is very close in spirit
to that of Lemma 2. 

\begin{lemma}\label{ramified}
For any prime $\ell$, we have
$$v_\ell(P(n_1, n_2)) = O(n_2 - n_1),$$
where the $O$-constant depends on $p$ only.
\end{lemma}
\begin{proof}
Let $r_1, \dots, r_k$ be the roots of $p$ in $\overline{{\mathbb
Q}}_{\ell}$, repeated according to their multiplicities. We have
$$v_{\ell}(p(x)) = v_{\ell}(\lc(p)) + \sum_{j=1}^k v_{\ell}(x - r_j).$$

Hence, 
\begin{eqnarray*}
v_{\ell}(P(n_1, n_2)) & = & (n_2 - n_1) v_{\ell}(\lc(p)) + \sum_{x=n_1}^{n_2-1} \sum_{j=1}^k v_{\ell}(x - r_j)\\
& = & \sum_{j=1}^k \sum_{x=n_1}^{n_2-1} v_{\ell}(x - r_j) + O(n_2 - n_1).\\
\end{eqnarray*}

Now, as in the proof of Lemmata 1-2, for each $j$, the number of $x \in
[n_1, n_2[$
such that $x - r_j = 0 \bmod \ell^i$ is $O((n_2 - n_1)/\ell^i)$. Hence, 
$$\sum_{x=n_1}^{n_2-1} v_{\ell}(x - r_j) = O\left(\frac{n_2 - n_1}{\ell - 1}\right),$$
from which our claim follows. 
\end{proof}

We can now state our main theorem regarding the size of the fraction
$T/Q$ in reduced form:
\begin{theorem}\label{bound-lval-gcd}
For $\ell$ a prime not dividing $\Delta(pq)$, one has
$$v_{\ell}(\gcd(T(n_1, n_2), Q(n_1, n_2))) \geq 
\frac
{\min (\rho_\ell(p), \rho_\ell(q))}
{\ell - 1}(n_2 - n_1)
+ 
O\left(\frac{\log n_2}{\log \ell}\right).$$
\end{theorem}
\begin{proof}
For $0 \leq n_1 < k < n_2$,
put $\tau(n_1, k, n_2) = a(k) P(n_1, k) Q(k, n_2)$. 
Then we have
$$T(n_1, n_2) = \sum_{n=n_1}^{n_2 - 1} \tau(n_1, k, n_2).$$

Applying Lemma~\ref{lvaluation}, we see that
$$v_\ell(\tau(n_1, k, n_2)) = v_\ell(a(k)) + \frac{\rho_\ell(p)}{\ell -
1}(k-n_1) + \frac{\rho_\ell(q)}{\ell - 1} (n_2-k) + O\left(\frac{\log n_2}{\log \ell}\right).$$

As such, 
$$v_{\ell}(T(n_1, n_2)) \geq 
\min_{n_1 < k < n_2} v_\ell(\tau(n_1, k, n_2)) \geq 
\frac
{\min (\rho_\ell(p), \rho_\ell(q))}
{\ell - 1}(n_2 - n_1)
+ 
O\left(\frac{\log n_2}{\log \ell}\right).$$

Joined with Lemma~\ref{lvaluation} for $v_{\ell}(Q(n_1, n_2))$, this
gives the desired statement.
%
\end{proof}
Note also that it is clear from the proof that this lower bound is 
generically sharp.

\begin{coro}
The following holds:
\begin{itemize}
\item If $p(n)$ and $q(n)$ have only linear irreducible factors,
the fraction $T(n_1, n_2)/Q(n_1, n_2)$
in reduced form has size $O(\min(n_2, (n_2 - n_1) \log n_2))$. 
\item
Otherwise, heuristically, as soon as $n_1 = o(n_2)$,
it is of size $\Theta(n_2 \log n_2)$.
\end{itemize}
\end{coro}
\begin{proof}
In the first case, the second part of the $O$-estimate is the trivial
estimate for the size of $Q(n_1, n_2)$. Since
$S(n_1,n_2)=\frac{T(n_1,n_2)}{Q(n_1,n_2)}$ is a partial sum of a
converging series, we have $T(n_1, n_2) = O(Q(n_1, n_2))$.  Therefore
this second part holds also for the size of the reduced fraction. 

The fact that $T(n_1, n_2) = O(Q(n_1, n_2))$ also implies that the size
of the reduced fraction is 
$\log \frac{Q(n_1, n_2)}{\gcd(T(n_1, n_2), Q(n_1, n_2))} + O(1)$. 

Now, we have
$$
\log \frac{Q(n_1, n_2)}{\gcd(T(n_1, n_2), Q(n_1, n_2))} 
 =  \sum_{\ell \text{ prime}} [v_\ell(Q(n_1, n_2)) - v_\ell(\gcd(T(n_1, n_2), Q(n_1, n_2)))] \log \ell.$$

However, since $q = \prod_{i=1}^k q_i^{e_i}$, we can discard in this
sum primes larger than $C(q) n_2$ for some constant $C(q)$ such
that $|q_i(x)| \leq C(q) n_2$ for all $i, x\in[n_1, n_2-1]$ (recall that
$q$ has only linear factors).
Further, the finitely many primes dividing the discriminant $\Delta$ of a prime
factor of $pq$ contribute for $O(n_2 - n_1)$ by
Lemma~\ref{ramified}. The logarithmic height therefore rewrites as:

$$
 \sum_{\substack{\ell \leq C(q) n_2, \, \ell \text{
 prime}\\(\ell,\Delta(pq))=1}} \left(\frac{n_2 - n_1}{\ell-1}
 [\rho_\ell(q) - \min\left(\rho_\ell(p),
 \rho_\ell(q)\right)]\log \ell + O(\log n_2)\right) + O(n_2-n_1).
$$
Under our assumptions, we have $\rho_\ell(p)=\rho_\ell(q)=\deg p=\deg q$,
hence this is also
$$\sum_{\substack{\ell \leq C(q)n_2, \, \ell \text{ prime}\\(\ell,\Delta(pq))=1}} O(\log n_2) + O(n_2-n_1) = O(n_2).$$

We now turn to the case where 
$q$ has irreducible factors of degree greater than $1$.
In this situation, it is preferable to compute the size of the
reduced fraction by subtracting the log of the
gcd to the asymptotic value of 
$\log Q(n_1, n_2) = (n_2 \log n_2 - n_1 \log n_1)\deg q + O(n_2 - n_1) = 
(n_2 - n_1)\deg q \log n_2  + O(n_2 - n_1)$, since
\begin{eqnarray*}
n_2 \log n_2 - n_1 \log n_1 & = & (n_2 - n_1) \log n_2 + n_1 \log n_2 / n_1 \\
&  & (n_2 - n_1) \log n_2 + O(n_2 - n_1).
\end{eqnarray*}

Again, write 
$$
\gcd(T(n_1, n_2), Q(n_1, n_2)) = \sum_{\ell \text{ prime}} v_\ell(\gcd(T(n_1, n_2), Q(n_1, n_2))) \log \ell.
$$

Heuristically, almost only primes of the order of magnitude of at most
$n_2 - n_1$ should appear both in $T$ and in $Q$.
Joined with the heuristic remark following Theorem~\ref{bound-lval-gcd},
this means that we expect the size of the gcd to be of the order of
$$(n_2 - n_1)\sum_{\ell \leq n_2 - n_1, \, \ell \text{ prime}} 
\frac{\min(\rho_\ell(p), \rho_\ell(q))}{\ell - 1}\log \ell + O(n_2).$$


\def\al{(\ell, {\mathbb K}/{\mathbb Q})} 
Recall $\deg p = \deg q$, and let ${\mathbb K}$
be the splitting field of $pq$. Denote by $\P$ the set of primes
$\ell$ such that $\rho_\ell(p) = \rho_{\ell}(q) = \deg (p) (= \deg (q))$.

The size of the gcd is 
\begin{eqnarray*}
& \leq & (n_2 - n_1) \sum_{\substack{\ell \leq n_2 - n_1, \,
\ell \text{ prime}\\\ell\not \in \P}} \frac{\deg q-1}{\ell -
1}\log \ell\\
 && + (n_2 - n_1) \sum_{\substack{\ell \leq n_2 - n_1, \, \ell \text{
 prime}\\\ell\in \P}} \frac{\deg q}{\ell - 1}\log\ell + O(n_2)\\
& = & (n_2 - n_1) \log(n_2 - n_1) (\deg q - 1) + (n_2 - n_1)
\sum_{\substack{\ell \leq n_2 - n_1, \, \ell\text{ prime}\\\ell\in \P}} \frac{\log \ell}{\ell - 1} + O(n_2). 
\end{eqnarray*}

\def\gl{{\mathfrak l}}
\def\agl{[\gl, {\mathbb K}/{\mathbb Q}]}
\def\id{\mathop{\mathrm{id}\relax}}
The last sum over primes evaluates to
\begin{eqnarray*}
\sum_{\substack{\ell \leq n_2 - n_1, \, \ell\text{ prime}\\\ell\in \P}}
\frac{\log \ell}{\ell - 1} & = & 
\sum_{\substack{\ell \leq n_2 - n_1, \, \ell\text{ prime}\\\ell\in \P}}
\frac{\log \ell}{\ell} + O(1),\\
\end{eqnarray*}
which, by classical analytic number theory arguments ---notice that 
the primes of $\P$ are exactly those for which the Artin
symbol $(\ell, {\mathbb K}/{\mathbb Q})$ equals $1$---, is 
$$\frac{\log (n_2 - n_1)}{[{\mathbb K}:{\mathbb Q}]} + O(1).$$
Thus, the size of the gcd is at most 
$$(n_2 - n_1) \log(n_2 - n_1) \left(\deg q - 1 + \frac{1}{[{\mathbb K}:{\mathbb Q}]}\right) + O(n_2).$$

Hence, under this heuristic, we see that the size of the reduced
fraction can be $O(n_2)$ when $n_1 = o(n_2)$ only if $[{\mathbb K}:{\mathbb Q}] = 1$,
which means that $p$ and $q$ are products of linear factors.
\end{proof}

\noindent
{\bf Remark.} In the proof of Corollary 1, we show that if 
$p$ and $q$ have linear factors only,
the size of the gcd is\footnote{Taking into
account only primes $\leq n_2 - n_1$, but any compensation occurring
for significantly larger primes is bound to be coincidental.} 
$(n_2 - n_1) \log (n_2 - n_1) \deg q + O(n_2)$. As long as $(n_2 - n_1)\log (n_2 -
n_1) = O(n_2)$, the size of the gcd is negligible with respect to the size of 
$Q$, which means that there is almost no compensation
between numerator and denominator of the fraction. Thus, compensations
start to appear in the evaluation tree only as soon as $n_2
- n_1$ is of the order of $n_2/\log n_2$.


\section{Using a Fully Factored Representation}
\label{sec:our-algo}

We extend in this section the ``partially factored representation'' of
\cite[Section 4]{ChZi00} to a ``fully factored representation'' for $P$,
$Q$, and $T$.

\subsection{Factored representation of integers}

We consider a set $B$ of primes. (In practice, it will consist of all
primes
needed to completely factor $p(n)$ and $q(n)$ up to $n=N-1$.)
A \emph{factored representation over $B$} of an integer $z$ is an
expression of the form:
\[ z=\prod_{p \in B} p^{\alpha_p} \cdot r, \]
where $\alpha_p \in {\mathbb N}$ and $r \in \Z$. The integer $z$ is
represented by the data $\left((\left\langle p,\alpha_p\right\rangle)_p,r\right)$. For efficiency
purposes, the representation skips primes $p$ such that $\alpha_p$ is zero.
Note that we do not impose that primes in $B$ do not divide the cofactor $r$, so
different factored representations may correspond to the same integer,
like $2^2 \cdot 3 \cdot 7$ and $2 \cdot 3 \cdot 14$ over $B = \{ 2, 3
\}$.  When the cofactor $r$ equals $1$, we have the (unique) \emph{fully
factored} representation of $z$.

\begin{figure}
\leftskip=0pt
\hrule
\vskip 4pt
{\small\obeylines
Algorithm \textbf{FastEval}($n_1$, $n_2$, $B$).
\tt
if ($n_1$ == $n_2 -1$) \{\qquad\qquad\qquad /* \emph{leaf computation} */
\qquad	$P$ = FullFactor($p(n_1)$, $B$);
\qquad	$Q$ = FullFactor($q(n_1)$, $B$);
\qquad	$T$ = FullFactor($p(n_1)$, $B$) $\cdot$ $a(n_1)$;
\} else \{
\qquad	$m$ = $\lfloor \frac{n_1+n_2}{2} \rceil$;
\qquad	$(P_1,Q_1,T_1)$ $\leftarrow$ FastEval($n_1$, $m$, $B$);
\qquad	$(P_2,Q_2,T_2)$ $\leftarrow$ FastEval($m$, $n_2$, $B$);
\qquad	$P$ = PartialMult($P_1$, $P_2$);
\qquad	$Q$ = PartialMult($Q_1$, $Q_2$);
\qquad	$T$ = PartialAdd(PartialMult($Q_2$, $T_1$), PartialMult($P_1$, $T_2$));
\}
}
\vskip 4pt
\hrule
\caption{Binary splitting using factored representation}
\label{pseudocode:fasteval}
\end{figure}

A binary splitting method using factored representations of integers can
be written as in algorithm \textbf{FastEval} (see
Fig.~\ref{pseudocode:fasteval})\footnote{ In algorithm \textbf{FastEval},
we do not factor the $a(n_1)$ term from $T$, which will not in general
share (additional) common primes with the denominator $Q$.}.  We need to
define the three operations FullFactor, PartialMult and PartialAdd.
FullFactor computes the full factorization of a given integer over the
factor base $B$, and is obtained by sieving (see below). Further, we
define:

\[ {\tt PartialMult}(\prod_{p \in B} p^{\alpha_p} \cdot r,
\prod_{p \in B} p^{\beta_p} \cdot s)
= \prod_{p \in B} p^{\alpha_p+\beta_p} \cdot (rs). \]

\[ {\tt PartialAdd}(\prod_{p \in B} p^{\alpha_p} \cdot r,
\prod_{p \in B} p^{\beta_p} \cdot s)
= \prod_{p \in B} p^{\gamma_p} \left( \prod_{p \in B}
p^{\alpha_p-\gamma_p}
\cdot r + \prod_{p \in B} p^{\beta_p-\gamma_p} s \right), \]
where $\gamma_p = {\rm min}(\alpha_p,\beta_p)$.

\subsection{Leaf computations}
We have $P(n,n+1) = p(n)$, $Q(n,n+1) = q(n)$, $T(n,n+1) = a(n) p(n)$.
The algorithm in Figure~\ref{pseudocode:fasteval} requires 
computing the fully factored representation of these quantities.
This corresponds to the leaves of the evaluation tree.

In order to expect an improvement from the use of the fully factored
representation, we must make sure that the gain is not offset by the complexity
of the leaf computations. In order to perform this step efficiently, we
use a standard window sieving method.

As mentioned in the introduction, we assume here that $p(n)$ and $q(n)$
are polynomials with linear factors only --- like in the case of
Formulae (\ref{pi3e}) or (\ref{zeta3}).
Without loss of generality, we illustrate our sieving
procedure with the computation of the fully factored representation of
the quantity $q(n)$ from Formula (\ref{zeta3}). This is equivalent to the
factorization of $2n+1$. The sieving produces simultaneously the
factorization of all the consecutive odd integers in a range $[2n_1+1,
2(n_1+W)+1[$, where $W$ is an arbitrary integer. We proceed as follows.
\begin{enumerate}
\item For each odd prime (or prime power) $\ell$ such that
$3\leq\ell<2N$, we compute the smallest value $i_\ell$ such that
$$2(n_1+i_\ell)+1\equiv0\mod\ell.$$
\item Sieve using the procedure in Figure~\ref{fig:sievingcode}.
\end{enumerate}
Besides this description, the important observation
is that the \emph{next} set of initialization values $i_{\ell}$ for the
computation of $q(n_1+W),\ldots,q(n_1 + 2W-1)$ does not have to be
computed: the code in Figure~\ref{fig:sievingcode} has already updated
these values correctly.

The translation of this scheme to other factorizations than that of
$2n+1$, as long as we stick to linear polynomials, is straightforward.

\begin{figure}
\leftskip=0pt
\hrule
\vskip 4pt
{\small\obeylines
Algorithm \textbf{Sieve}($n_1$, $n_1 + W - 1$)
\tt
factored\_representation $\tau$[$W$]
for all primes $\ell<2N$
\qquad	$i$ = $i_\ell$
\qquad	while ($i < W$) \{ %
		include $\ell$ in $\tau$[i] ; %
		$i$ = $i + \ell$ \}
\qquad	$i_\ell$ = $i - W$
\qquad	for all powers $\ell^k$ of $\ell$, with $\ell^k<2N$
\qquad	\qquad	$i$ = $i_{\ell^k}$
\qquad	\qquad	while ($i < W$) \{ %
increase by one the multiplicity of $\ell$ in $\tau$[i]; %
$i$ = $i + {\ell^k}$ \}
\qquad	\qquad	$i_{\ell^k}$ = $i - W$
}
\vskip 4pt
\hrule
\caption{Pseudo-code for sieving}
\label{fig:sievingcode}
\end{figure}

\section{Analysis of the algorithm}
\label{sec:analysis}

\subsection{Cost of sieving}

Because $p(n), q(n)$ are assumed to have linear factors only, their
prime divisors are bounded by $c n$ for some constant $c$,
thus all $p(n), q(n)$ for $n \leq N$ can be completely factored
over a set of $O(N/\log N)$ primes or prime powers.
Since the initialization of the sieve only has to be done once, the
computation of the $i_\ell$ values is trivial. In total, the sieving code
in Figure~\ref{fig:sievingcode} performs $O(N/\ell)$ sieve updates for
each prime (or prime power) $\ell$.  The number of times the sieve
procedure is called is $O(N/W)$, and each time all of the $O(N/\log N)$
primes or prime powers are scanned. This yields a time complexity for
sieving which is $O(N\log\log N + N^2/(W\log N))$. The space complexity
for sieving is at most $O(W\log N)$.  There is some freedom in the choice
of $W$, but it must clearly be between $O(N/(\log N)^4)$ and $O(N/\log
N)$, so that the time and space complexities remain below $O(d\log^3 d)$
and $O(d)$, respectively.

\subsection{The recursion}

The factor base $B$ consists of all possible prime divisors of
$p(n)$ or $q(n)$, which means $O(N/\log N)$ primes.
The integers $P$ and $Q$ are therefore always fully factored. Computing the
product $P(n_1,n_2) = P(n_1,m) P(m,n_2)$ thus just consists in
adding the prime exponents in the lists of factors. If the factored
representations of $P(n_1,m)$ and $P(m,n_2)$ have
respectively $l_1$ and $l_2$ elements, this can be done in
$O(l_1+l_2)$ 
operations.

At level $k$ --- the leaves corresponding to level $0$ --- the
values of $P$ or $Q$ are bounded by $O((N^{\deg p})^{2^k})$,
thus have $O(2^k \log N)$ bits.
On the other hand, let $l$ be the number of different prime factors
in the representation of $P$ (counted with multiplicities),
then $P \geq 2^l$, thus $P$ has $\Omega(l)$ bits.
It follows that at level $k$, we have $l = O(2^k \log N)$.
(We also have $l = O(\frac{N}{\log N})$ since there are that number
of primes
in the factor base $B$.)

The total cost of computing $P$ and $Q$ is thus bounded by
$\sum_{k=0}^{\log N} \frac{N}{2^k} (2^k \log N) = O(d \log^2 d)$.

\medskip

As concerns $T$, if we could prove that its non-factored part is
always
$\log N$ times smaller than the factored part, we would get a
complexity
of $O(M(d) \log d)$ for $T$.
Indeed, at level $k$, the non-factored part of $T$ would have
$O(2^k)$ bits, thus the cost of computing it would be $O(M(2^k))$,
since it is obtained from a sum of products
$T(n_1,m) Q(m,n_2) + P(n_1,m) T(m,n_2)$, where no cancellation
occurs.
Thus the total cost for the non-factored part would be
$\sum_{k=0}^{\log N}
\frac{N}{2^k} O(M(2^k)) = O(M(d) \log d)$.
(The analysis for the factored part is similar to that for $P$ and
$Q$.)

Unfortunately, the above property --- the non-factored part of $T$ is
$\log N$ times smaller than its factored part --- is only true near
the root
of the product tree, where common factors cancel between $P$ and $Q$.
Thus the computation of $T$ costs $O(M(d) \log^2 d)$ (unless we can
do better).

\section{Experimental Results}
\label{sec:experiments}

In this section we investigate the benefit of using the fully factored
representation for the purpose of computing the sum of series such
as~(\ref{pi3e}) or~(\ref{zeta3}).

Compared to the sieve table of the \texttt{gmp-chudnovsky} program
mentioned in Section~\ref{sec:algo-classic}, we address the problem of
the leaf computation in a different way. The sieving procedure described
in Section~\ref{sec:our-algo} keeps a space complexity of the order of
$O(d)$. We found that our sieving procedure was competitive with the
sieve table from the \texttt{gmp-chudnovsky} program.

The fully factored representation is an asset as soon as compensations
between prime factors start to appear. However, this is not encountered at
the very lowest levels of the computation tree, near the leaves. Quite
naturally, a cut-off level appears between the use of the factored
representation and the use of the fully expanded integer values. At the
lowest levels of the tree, we use the same approach as the
\texttt{gmp-chudnovsky} program. For $P$ and $Q$, both the expanded
integer and its factorization are kept. The integer component is dropped
above a certain height in the tree.  The remark concluding
Section~\ref{sec:analysis} suggests that the switch from one algorithm to
the other be done when $n_2-n_1$ is of the order of $\frac{n_2}{\log
n_2}$.  However the running time is the measurement here, and the precise
cut-off is chosen by trial and error.

Another implementation note concerns the PartialAdd operation mentioned
in Section~\ref{sec:our-algo}, and also the final expansion of $T$ and
$Q$ from the factored form to a flat integer. For this purpose, we use
the same kind of algorithm as mentioned in~\cite{Schonhage94} for the
computation of $n!$.
\medskip

\begin{table}
\begin{center}
\begin{tabular}{l|r|r|c}
\multicolumn{4}{c}{Opteron, 2.4Ghz} \\ \hline
digits & our code & \texttt{gmp-chudnovsky} & ratio\\
\hline
$2^{25}$&	60s	&61s		&0.98\\
$2^{26}$&	136s	&147s		&0.93\\
$2^{27}$&	322s	&352s		&0.91\\
$2^{28}$&	768s	&853s		&0.90\\
$2^{29}$&	1868s	&2059s		&0.91\\
$2^{30}$&	4654s	&5328s		&0.87\\
\omit\\[3pt]
\multicolumn{4}{c}{Pentium 4, 3Ghz} \\ \hline
digits & our code & \texttt{PiFast} & ratio\\
\hline
$28\times 10^6$ & 127s & 135s & 0.94\\
$40\times 10^6$ & 192s & 206s & 0.93\\
$57\times 10^6$ & 291s & 323s & 0.90\\
\end{tabular}
\end{center}
\caption{Comparison with the \texttt{gmp-chudnovsky} and \texttt{PiFast}
programs for computing digits of $\pi$.}
\label{timings}
\end{table}

We implemented our algorithm in C++ with the GMP and MPFR libraries
\cite{Granlund06a,FoHaLePeZi06}. We
modified the GMP library with an improved FFT multiplication
code~\cite{GaKrZi06z}. We compare our results with the two programs
mentioned in Section~\ref{sec:algo-classic}, which compute digits of
$\pi$ using Formula (\ref{pi3e}). For the purpose of comparison, we focus
on the time for the evaluation of the fraction $T/Q$.  Because our
program and the \texttt{gmp-chudnovsky} program share the GMP library as
a common backbone, we are convinced that this comparison gives the most
meaningful results.  Table~\ref{timings} gives the relative time spent in
the binary splitting process for our program and for the
\texttt{gmp-chudnovsky} program, measured on a 2.4Ghz Opteron CPU. The
gain seems to grow slightly with the number of digits computed.

Table~\ref{timings} also mentions timings of our program against the
\texttt{PiFast} program. The comparison has been made on a 3 GHz Pentium
4 CPU (hence the different timings). The lack of source code access for
\texttt{PiFast} mandates some caution for the interpretation of the
timings, since the operating system overhead seems to be included.
Nonetheless, it seems that our fully factored binary splitting provides a
growing improvement.

\smallskip

Finally, we used our program to establish a new record-size computation
for 2 billion decimal digits of $\zeta(3)$.  The series of
Formula (\ref{zeta3}) was evaluated up to $N=664385619$ terms.
The computation of $Q$ and $T$ was first spanned over $16$ distinct
2.4Ghz Opteron processors. The cumulative CPU time spent by these
processors to compute their fraction of the result was $20$ hours.
These results were gathered to form the final fraction $T/Q$ (in factored
form) on a single computer in $3$ hours. Converting the fraction $T$ and
$Q$ to integers took $18$ minutes.
The division took $53$ minutes,
and the decimal conversion took $2$ hours and $37$ minutes
(these timings are suboptimal, since GMP does not yet implement Newton's
division).

\smallskip
\noindent
\textsc{Error analysis.}
Using the inequality $\frac{n}{2n+1}<\frac12$ and a rough bound on
$a(n)$,
it can be shown that Formula (\ref{zeta3}) gives an error of at most
$2^{-10N+2\log_2 N + 2}$, which is less than $2^{-6643856129}$ here.
Using a precision of $p=6643856189$ bits, we converted $T$ and $Q$ to
floating-point numbers, divided both, and converted the binary quotient to
a decimal string of $2\cdot10^9+1$ digits, all those operations being made in
rounding to nearest mode with the MPFR library.
An error analysis yields
a maximal absolute error of $2^{1-p}$, which together with the above truncation
error gives a maximal absolute error of $10^{-1999999981}$.

\bigskip

\noindent
\textbf{Acknowledgement.} The authors thank Richard Brent for his clever
remarks on a first draft of this paper.

\nocite{Schonhage94}


\begin{thebibliography}{1}

\bibitem{Be04}
{\sc Bernstein, D.~J.}
\newblock Fast multiplication and its applications.
\newblock \url{http://cr.yp.to/papers.html#multapps}, 47 pages, 2004.

\bibitem{BoBo87}
{\sc Borwein, J. and Borwein, P.}
\newblock Pi and the AGM.
\newblock John Wiley and Sons, 1987.

\bibitem{BoBrCr00}
{\sc Borwein, J. and Bradley, D. and Crandall, R.}
\newblock Computational stratgies for the Riemann zeta function.
\newblock {\em Journal of Computational and Applied Mathematics 121}
(2000), 247--296.

\bibitem{Br76}
{\sc Brent, R.}
\newblock Fast multiple-precision evaluation of elementary functions.
\newblock {\em Journal of the ACM 23}, 2 (1976), 242--251.

\bibitem{ChGeKiZi05}
{\sc Cheng, H., Gergel, B., Kim, E., and Zima, E.}
\newblock Space-efficient evaluation of hypergeometric series.
\newblock {\em SIGSAM Bulletin, Communications in Computer Algebra
39}, 2
  (2005), 41--52.

\bibitem{ChZi00}
{\sc Cheng, H., and Zima, E.}
\newblock On accelerated methods to evaluate sums of produts of
rational numbers.
\newblock In {\em Proceedings of ISSAC'00\/} (2000), pp.~54--61.

\bibitem{FoHaLePeZi06}
{\sc Fousse, L., Hanrot, G., Lef{\`e}vre, V., P{\'e}lissier, P., and
  Zimmermann, P.}
\newblock {MPFR}: A multiple-precision binary floating-point library with
  correct rounding.
\newblock {\em ACM Trans. Math. Softw. 33}, 2 (2007).

\bibitem{GaKrZi06z}
{\sc Gaudry, P., Kruppa, A., and Zimmermann, P.}
\newblock A GMP-based implementation of Sch\"onhage-Strassen's large
integer multiplication algorithm.
\newblock Submitted, 2007.

\bibitem{Go90}
{\sc Gosper, R.}
\newblock Strip mining in the abandoned orefields of nineteenth
century mathematics.
\newblock {\em Computers in Mathematics} (1990), pp.~261--284.

\bibitem{GoSe01}
{\sc Gourdon, X., and Sebah, P.}
\newblock Binary splitting method.
\url{http://numbers.computation.free.fr/Constants/Algorithms/splitting.html},
Jan. 2001.

\bibitem{Granlund06a}
{\em {GMP}: {T}he {GNU} {M}ultiple {P}recision {A}rithmetic {L}ibrary},
  4.2.1~ed., 2006.
\newblock \url{http://www.swox.se/gmp/}.

\bibitem{HaPa98}
{\sc Haible, B., and Papanikolaou, T.}
\newblock Fast multiprecision evaluation of series of rational numbers.
\newblock Algorithmic Number Theory, Third International Symposium, ANTS-III,
\newblock volume 1423 of {\em Lecture Notes in Computer Science}, Springer,
  1998.

\bibitem{Karatsuba97}
{\sc Karatsuba, E.~A.}
\newblock Fast evaluation of hypergeometric functions by {FEE}.
\newblock In {\em Proceedings of Computational Methods and Function
Theory
  (CMFT'97)\/} (1997), N.~Papamichael, S.~Ruscheweyh, and E.~B. Saff,
Eds.,
  World Scientific Publishing, pp.~303--314.

\bibitem{Schonhage94}
{\sc Sch{\"o}nhage, A., Grotefeld, A. F.~W., and Vetter, E.}
\newblock {\em Fast Algorithms, A Multitape {T}uring Machine
Implementation}.
\newblock BI-Wissenschaftsverlag, 1994.

\bibitem{ScSt71}
{\sc Sch\"onhage, A. and Strassen, V.}
\newblock Schnelle Multiplikation gro\ss er Zahlen.
\newblock {\em Computing 7} (1971), 281--292.

\bibitem{pan-wang:2003}
{\sc Wang, X. and Pan, V.~Y.}
\newblock Acceleration of Euclidean algorithm and rational number
  reconstruction.
\newblock {\em SIAM Journal on Computing}, 32(2):548--556, 2003.

\bibitem{Xue02}
{\sc Xue, H.}
\newblock \texttt{gmp-chudnovsky.c} code for computing digits of $\pi$
using the Gnu MP library.
\newblock
Available at \url{http://www.swox.com/gmp/pi-with-gmp.html}, 2002.

\end{thebibliography}
\end{document}